\input phyzzx.tex
\tolerance=1000
\voffset=-0.0cm
\hoffset=0.7cm
\sequentialequations
\def\rl{\rightline}

\def\t1{{\tilde 1}}

\def\t{\theta}

\REF{\HOL}{G. 't Hooft, gr-qc/9310026; L. Susskind, J. Math. Phys. {\bf 36} (1995) 6377, hep-th/9409089.}
\REF{\RAP}{R. Bousso, JHEP {\bf 9907} (1999) 004, hep-th/9905177; JHEP {\bf 9906} (1999) 028, hep-th/9906022; JHEP {\bf 0104} (2001) 035, hep-th/0012052.}
\REF{\GH}{G. W. Gibbons and S. W. Hawking, Phys. Rev. {\bf D15} (1977) 2738.}
\REF{\MAL}{J. Maldacena and A. Strominger, JHEP {\bf 9802} (1998) 014, gr-qc/9801096.}
\REF{\HMS}{S. Hawking, J. Maldacena and A. Strominger, JHEP {\bf 0105} (2001), hep-th/0002145.}
\REF{\WIT}{E. Witten, hep-th/0106109.}
\REF{\LIN}{F. Lin and Y. Wu, Phys. Lett. {\bf B453} (1999) 222, hep-th/9901147.}
\REF{\STR}{A. Strominger, JHEP {\bf 0110} (22001) 034; hep-th/0106113.}
\REF{\SSV}{M. Spradlin, A. Strominger and A. Volovich, hep-th/0110007.}
\REF{\CON}{J. D. Brown and M. Henneaux, Commun. Math. Phys. {\bf 104} (1986) 207.}
\REF{\VER}{E. Verlinde, hep-th/0008140.}
\REF{\BEK}{J. Bekenstein, Lett. Nuov. Cimento {\bf 4} (1972) 737; Phys Rev. {\bf D7} (1973) 2333; Phys. Rev. {\bf D9} (1974) 3292.}
\REF{\HAW}{S. Hawking, Nature {\bf 248} (1974) 30; Comm. Math. Phys. {\bf 43} (1975) 199.}
\REF{\CAR}{J. L. Cardy, Nucl. Phys. {\bf B270} (1986) 186.}
\REF{\DAN}{U. H. Danielsson, hep-th/0110265.}
\REF{\KLE}{D. Klemm, hep-th/0106247.}
\REF{\NO}{S. Nojiri and S. D. Odintsov, hep-th/0106191; hep-th/0107134.}
\REF{\DES}{E. Halyo, hep-th/0107169.}
\REF{\INF}{A. Strominger, hep-th/0110087.}
\REF{\SIT}{T. Shiromizu, D. Ita and T. Torii, hep-th/0109057.}
\REF{\CAI}{R. G. Cai, hep-th/0111093.}
\REF{\MED}{A. J. M. Medved, hep-th/0111182; hep-th/0111238.}
\REF{\OGU}{S. Ogushi, hep-th/0111008.}
\REF{\BBM}{V. Balasubramanian, J. de Boer and D. Minic, hep-th/0110108.}
\REF{\PET}{A. C. Petkou and G. Siopsis, hep-th/0111085.}
\REF{\MAN}{A. M. Ghezelbash and R. B. Mann, hep-th/0111217.}
\REF{\CVE}{M. Cvetic, S. Nojiri and S. D. Odintsov, hep-th/0112045.}
\REF{\BK}{V. Balasubramanian and P. Kraus, Comm. Math. Phys. {\bf 208} (1999) 413, hep-th/9902121.}
\REF{\NAR}{H. Nariai, Sci. Rep. Tohuku Univ. {\bf 34} (1950) 160; {\it ibid.} {\bf 35} (1951) 62.}
\REF{\BY}{J. D. Brown and J. W. York, Phys. Rev {\bf D47} (1993) 1407.}
\REF{\LEN}{L. Susskind, hep-th/9309145.}
\REF{\SBH}{E. Halyo, A. Rajaraman and L. Susskind, Phys. Lett. {\bf B392} (1997) 319, hep-th/9605112.}
\REF{\HRS}{E. Halyo, B. Kol, A. Rajaraman and L. Susskind, Phys. Lett. {\bf B401} (1997) 15, hep-th/9609075.}
\REF{\EDI}{E. Halyo, Int. Journ. Mod. Phys. {\bf A14} (1999) 3831, hep-th/9610068; Mod. Phys. Lett. {\bf A13} (1998), hep-th/9611175.}
\REF{\UNI}{E. Halyo, hep-th/0108167.}

\singlespace
\rl{SU-ITP-01-53}
\rl{hep-ph/0112093}
\rl{\today}
\pagenumber=0
\normalspace
\medskip
\bigskip
\titlestyle{\bf{On the Cardy--Verlinde Formula and the de Sitter/CFT Correspondence}}
\smallskip
\author{ Edi Halyo{\footnote*{e--mail address: vhalyo@stanford.edu}}}
\smallskip
\centerline {Department of Physics}
\centerline{Stanford University}
\centerline {Stanford, CA 94305}
\centerline{and}
\centerline{California Institute for Physics and Astrophysics}
\centerline{366 Cambridge St.}
\centerline{Palo Alto, CA 94306}
\smallskip
\vskip 2 cm
\titlestyle{\bf ABSTRACT}

We derive the Cardy--Verlinde entropy formula for the field theory that lives on the boundary of an asymptotically de Sitter space with a black hole.
The boundary theory which is not conformal has a monotonic $C$--function defined by the Casimir energy. The instability of the space due to
Hawking radiation from the black hole corresponds to an RG flow from the IR to the UV during which $C$ increases. The endpoint of black hole evaporation
is de Sitter space which is described by a conformal theory at the UV fixed point of the RG flow.

\singlespace
\vskip 0.5cm
\endpage
\normalspace

\centerline{\bf 1. Introduction}
\medskip

Holography is believed to be one of the fundamental principles of the true quantum theory of gravity[\HOL,\RAP]. An explicitly calculable example of holography
is the much--studied AdS/CFT correspondence. Unfortunately, it seems that we live in a universe with a positive cosmological constant which will look like de Sitter
space--time in the far future. Therefore, we should try to understand quantum gravity or string theory in de Sitter space preferably in a holographic way. Of course,
physics in de Sitter space is interesting even without its connection to the real world; de Sitter entropy and temperature have always been mysterious aspects
of quantum gravity[\GH]. (For previous work on this subject see [\MAL-\LIN].)

Recently, a holographic duality between de Sitter space--time and an Euclidean CFT living on the de Sitter boundary was conjectured[\STR,\SSV].
This dS/CFT correspondence is similar
to the well--known AdS/CFT correspondence. The main supporting evidence for the conjecture seems to be the asymptotic conformal symmetry on the boundary (as in [\CON]) and the
behavior of the boundary correlation functions. However, there are some important differences between the AdS/CFT and the dS/CFT dualities. For example, the CFT dual to
de Sitter space is Euclidean and not unitary. This makes the interpretation of concepts such as energy, temperature, central charge etc. quite difficult.
On the other hand, if the CFT truly describes the physics of de Sitter space, then it has to account for the nonzero entropy and temperature of de Sitter
(or asymptotically de Sitter) space.

Using the AdS/CFT duality, it was shown that the entropy of a CFT on $S^d$ (dual to an AdS black hole background) is
given by the Cardy--Verlinde formula[\VER]
$$S={{4 \pi R} \over {(d-2)}} \sqrt{E_E E_C}  \eqno(1)$$
where $E_E$ and $E_C$ are the extensive and Casimir parts of the CFT energy and $R$ is the radius of $S^d$.
The Cardy--Verlinde formula reproduces the Bekenstein--Hawking entropy [\BEK,\HAW] of the black hole in the bulk.
With a simple identification between $E_E$, $E_C$ and $L_0$, $c$
eq. (1) becomes the well--known Cardy formula for two dimensional (unitary) CFTs[\CAR].
It is surprising that the AdS/CFT correspondence implies that the formula applies to higher dimensional CFTs as well.
In ref. [\DAN] it was conjectured that a very similar formula (with a sign change) describes the entropy of a Euclidean CFT which is dual to asymptotically de Sitter space. Then,
with a proper definition of de Sitter energy, the modified Cardy--Verlinde formula accounts for the entropy of the cosmological horizon. The validity of the Cardy--Verlinde
entropy formula is even more surprising in the de Sitter case because the boundary CFT is not unitary. On the other hand, there seems to be some circumstantial evidence for
both the dS/CFT correspondence and the Cardy--Verlinde formula[\STR,\SSV,\DAN-\CVE].

In this paper, we derive the Cardy--Verlinde formula for asymptotically de Sitter spaces using the classical metric, the asymptotic conformal symmetry and the Bekenstein--Hawking
formula. Thus, if we take the Bekenstein--Hawking entropy as a fundamental feature of quantum gravity we are led to the (modified) Cardy--Verlinde entropy
formula for a CFT in $d$ dimensions.
The reason for the existence of such a formula for CFTs in more than two dimensions which are not unitary is not clear.
We then show that there is an effective boundary theory dual to asymptotically de Sitter space with formulas for energy and
entropy similar to those of two dimensional CFTs.
This theory has a monotonic $C$--function which is defined by the Casimir energy. We show that
$C$ is a nondecreasing function of the energy on the boundary. For an asymptotically de Sitter
space (with a black hole) $C<c_{dS}$; therefore, the boundary theory is not conformal, i.e. it sits away from the fixed point with $c_{dS}$.
The existence of a modified Cardy formula for entropy is even more mysterious in this case since the boundary theory is neither unitary nor conformal.
We interpret the instability in bulk time i.e. the Hawking radiation from the black hole as an RG flow of the boundary theory from the IR to the UV.
As time passes the black hole evaporates and its mass decerases.
This corresponds on the boundary theory to an RG flow from the IR to the UV during which $C$ increases. When the black hole completely evaporates and all the radiation crosses
the cosmological horizon, $M \to 0$ and $C \to c_{dS}$.
This final state which is the pure de Sitter space is stable and corresponds on the boundary to a CFT (at a UV fixed point).
Thus, the boundary description of Hawking radiation is integrating in
degrees of freedom in a flow towards the UV. $C$ is also bounded from below due to the existence of an upper bound on the black hole mass. This Nariai black hole
corresponds to an unstable UV fixed point of the boundary theory.
For asymptotically $dS_3$ spaces with mass $M$ we find that $C=c_{dS}$ for any $M$. The absence of an RG run agrees with the fact that in three
dimensions $M$ describes not a black hole but a pointlike mass with no horizon; there is no Hawking radiation in this case.

The paper is organized as follows. In section 2 we give the derivation of the Cardy--Verlinde entropy formula from the AdS/CFT correspondence. In section 3 we repeat the
same exercise for the dS/CFT correspondence. In section 4 we describe the effective CFT and how it is related to Hawking radiation from a black hole in asymptotically de Sitter
space. Section 5 includes a discussion of our results and our conclusions.

\medskip
\centerline{\bf 2. The Cardy--Verlinde Formula from the AdS/CFT Correspondence}
\medskip

In this section we derive the Cardy--Verlinde formula for the entropy of a CFT on $S^d$ using the AdS/CFT correspondence. This was implicitly done in ref. [\VER]; here we review
it in order to show (in the next section) that exactly the same exercise can be repeated for the dS/CFT correspondence.

Consider a large black hole (i.e. with $R>>L$) in $AdS_d$ with the metric
$$ds^2=-(1+{r^2 \over L^2}-{{2G_d M} \over r^{d-3}})dt^2+(1+{r^2 \over L^2}-{{2G_d M} \over r^{d-3}})^{-1}dr^2+r^2 d^2 \Omega_{d-2} \eqno(2)$$
where $L^{-2}=\Lambda$, $G_d$ is the Newton's constant and $M$ is the mass parameter of the black hole.
The position of the horizon is given by $R$ which solves
$$1+{R^2 \over L^2}-{{2G_d M} \over R^{d-3}}=0 \eqno(3)$$
This can be used as a definition of the black hole mass parameter ($M=M_1+M_2$)
$$M={R^{d-3} \over 2G_d}+{R^{d-1} \over {2G_d L^2}} \eqno(4)$$

The real quantity we need is the energy of the black hole as an excitation in $AdS_d$.
In principle the $AdS_d$ vacuum may also have nonzero energy. In fact in ref. [\BK] it has been shown
that the $AdS_d$ vacuum for $d$ odd has nonzero energy due to the contribution from the anomalous Casimir effect. (For even $d$ the anomalous Casimir energy vanishes.)
Thus, the total energy in $AdS_d$ can be written as $E_{AdS}=E_{C,A}+E_{BH}$.
However, we are only interested in the excitation energy of the black hole and therefore we will
subtract the contribution $E_{C,A}$ of the $AdS_d$ vacuum.
Then, from the calculation of the Brown--York tensor [\BY] the energy of the black hole
in $AdS_d$  is found to be
$$E_{AdS}={(d-2) \over {8 \pi}} M \eqno(5) $$
Consider now the CFT on the boundary of $AdS_d$, (at $r>>L$) which is an $S^{d-2}$ of radius $r$. Due to the
conformal symmetry of the boundary (the boundary metric is fixed up to conformal transformations) we can rescale the boundary coordinates so that the $S^{d-2}$ has
the same radius as the black hole, $R>>L$. From the metric in eq. (2) we see that the energy of the boundary CFT is redshifted by $L/R$ comprared to $E_{AdS}$.
Using the connection between the central charge of the CFT and Newton's constant $c=3L^{d-2}/G_d$ we get
$$E_{CFT}={c (d-2) \over {48 \pi}} {V \over L^{d-1}} (1+{L^2 \over R^2}) \eqno(6)$$
where $V=R^{d-2}$ is the volume of the boundary. The energy of the CFT $E_{CFT}=E_E+E_C$ has two contributions. The first term, $E_E$, is the usual extensive term
which describes the energy of the $(d-2)$--dimensional CFT gas.
The second term, $E_C$, is sub--extensive and gives the Casimir energy of the CFT on $S^{d-2}$.
The Hawking temperature of the black hole is
$$T_H={R \over {4 \pi L^2}}((d-1)+(d-3){L^2 \over R^2}) \eqno(7)$$
Therefore, the boundary CFT is at a temperature given by $T_{CFT}=(L/R)T_H$ due to the redshift.
Using
$$T_{CFT}=\left(\partial E_{CFT} \over \partial S \right)_V \eqno(8)$$
we obtain the entropy of the CFT
$$S_{CFT}={c \over 12} {V \over L^{d-2}}={4 \pi \over {(d-2)}} R \sqrt{E_E E_C} \eqno(9)$$
which is the Cardy--Verlinde formula for the entropy of a CFT on $S^{d-2}$.
It is easy to see that eq. (9) reproduces the Bekenstein--Hawking entropy[\BEK,\HAW]
$$S_{CFT}=S_{BH}={A \over {4G_d}} \eqno(10)$$
Of course, using the Hawking temperature is equivalent to using the Bekenstein--Hawking entropy formula. We stress that this is not a derivation of the
Bekenstein--Hawking formula but a validation of eq. (9) assuming that eq. (10) a fundamental component of any quantum theory of gravity.

We found that a (large) black hole in $AdS_d$ is holographically described by a thermal state of the boundary CFT with a central charge given by the bulk parameters
$L$ and $G_d$. The conformal symmetry of $AdS_d$ is broken by the black hole; in the boundary CFT this corresponds to the nonzero temperature $T_{CFT}$ (or radius of the boundary).
This black hole is stable and does not radiate which means that the boundary CFT is at thermal equilibrium.

The above derivation of the Cardy--Verlinde formula used three pieces of input. The first is the metric in eq. (2) which gives the equation
for the position of the black hole horizon (which is the equation for the mass of the black hole). The metric also determines the redshift between $E_{AdS}$ and $E_{CFT}$.
The second is the conformal symmetry of the boundary of de Sitter space
which allowed us to rescale the coordinates so that the boundary has a radius equal to that of the
black hole. The third is the Hawking temperature of the black hole or equivalently the Bekenstein--Hawking entropy which should be valid in quantum gravity.

\medskip
\centerline{\bf 3. The Cardy--Verlinde Formula and the dS/CFT Correspondence}
\medskip

It seems that the same three pieces of input used in the above derivation of
the Cardy-Verlinde formula are also present in the case of asymptotically de Sitter space. Since the metric for a black hole in $dS_d$ is very similar
(to that in eq. (2) with only a sign change) we expect that the definition of mass and the
energy redshift will also be similar to the $AdS_d$ case. Moreover, there is an asymptotic Euclidean
conformal symmetry on the boundary which can be used to rescale the coordinates.
Therefore, we can repeat the steps in the previous section for the dS/CFT correspondence and derive
the modified Cardy--Verlinde formula.

We now consider a $d$--dimensional asymptotically de Sitter space, i.e. a black hole in $dS_d$ described by the metric
$$ds^2=-(1-{r^2 \over L^2}-{{2G_d M} \over r^{d-3}})dt^2+(1+{r^2 \over L^2}-{{2G_d M} \over r^{d-3}})^{-1}dr^2+r^2 d^2 \Omega_{d-2} \eqno(11)$$
where $L^{-2}=\Lambda$.
The cosmological horizon is at $r=R<L$ which is given by
$$1-{R^2 \over L^2}-{{2G_d M} \over R^{d-3}}=0 \eqno(12)$$
As before this can be used as a definition of the black hole mass parameter ($M=M_1-M_2$)
$$M={R^{d-3} \over 2G_d}-{R^{d-1} \over {2G_d L^2}} \eqno(13)$$

As in the $AdS_d$ case the real quantity we need is the energy of the black hole as an excitation in $dS_d$.
In principle the $dS_d$ vacuum may also have nonzero energy. Recently, in ref. [\BBM] it was shown
that (in a manner very similar to the $AdS_d$ case) the $dS_d$ vacuum for $d$ odd
has nonzero positive energy due to the contribution from the anomalous
Casimir effect (whereas for even $d$ the energy vanishes). Thus the total energy in $dS_d$ can be written as $E_{dS}=E_{C,A}+E_{BH}$.
However, as before, we are only interested in the excitation energy of the black hole and therefore we will
subtract the contribution $E_{C,A}$ of the pure $dS_d$ vacuum.

Asymptotically de Sitter space in $d$ dimensions has two space--like boundaries for $t \to \infty$ and $t \to -\infty$ denoted by $I^+$ and $I^-$ respectively.
It has been shown from the calculation of the Brown--York tensor (on the future boundary $I^+$) that the energy of $dS_d$ with the black hole is
$$E_{dS}=-{(d-2) \over {8 \pi}} M \eqno(14) $$
The minus sign arises if we take $I^+$ as the boundary of $dS_d$ (for $I^-$ the sign is positive) which is a convention. Then, as the black hole mass parameter
$M$ increases both the entropy of the cosmological horizon and the total energy of the space decrease. Note that with this convention the energy of the space is negative
and it vanishes only for pure $dS_d$.

Consider now the Euclidean CFT on $I^+$, (at $r>>L$ which is way beyond the cosmological horizon) which is an
$S^{d-2}$ of radius $r$.
Note that for $r>R$, $g_{tt}>0$ and $g_{rr}<0$ so in this region $r$ is time--like and $t$ is space--like. For $r>>R$ the metric in eq. (11) becomes
$$ds^2=({r^2 \over L^2})dt^2-({L^2 \over r^2})dr^2+r^2 d^2 \Omega_{d-2} \eqno(15)$$
We see that the boundary at $r>>R$ is Euclidean and given by $R \times S^{d-2}$.
Due to the (Euclidean) conformal symmetry of the boundary we can
again rescale the boundary coordinates so that the $S^{d-2}$ has
the same radius as the cosmological horizon, $R<L$. From eq. (15) we see that the energy of the boundary theory should be redshifted by
$L/R$ comprared to $E_{dS}$. Using $c=3L^{d-2}/G_d$ we get
$$E_{CFT}={c (d-2) \over {48 \pi}} {V \over L^{d-1}} (1-{L^2 \over R^2}) \eqno(16)$$
$E_{CFT}=E_E-E_C$ has two contributions. The first term, $E_E$,  which is positive is the usual extensive term which describes the energy of the CFT gas.
The second term,$-E_C$, which is negative is sub--extensive and gives the Casimir energy of the
Euclidean CFT on $S^{d-2}$.

In this case, the temperature of asymptotically de Sitter space (or the cosmological horizon) is
$$T_{dS}={R \over {4 \pi L^2}}((d-1)-(d-3){L^2 \over R^2}) \eqno(17)$$
The temperature of the CFT is given by $T_{CFT}=(L/R)T_{dS}$. Using
$$T_{CFT}=\left(\partial E_{CFT} \over \partial S \right)_V \eqno(18)$$
we find the entropy of de Sitter space to be
$$S_{CFT}={4 \pi \over {(d-2)}} R \sqrt{E_E |E_C|} \eqno(19)$$
This is the modified Cardy--Verlinde formula for asymptotically de Sitter spaces which reproduces the Bekenstein--Hawking entropy of the cosmological horizon.

The Cardy--Verlinde formula in eq. (19) holds for both pure and asymptotically de Sitter spaces. For example, consider, $dS_d$ which is described by a CFT on the boundary.
In this case, the cosmological horizon is at $R=L$ and we find $E_E=E_C$ and $E_{CFT}=0$ (since there is no black hole). The redshift factor is one and we find
$T_{CFT}=1/2 \pi L$. The entropy of the space becomes $S_{CFT}=R^{d-2}/4 G_d$ as expected.
We can write the Cardy--Verlinde formula in eq. (19) in a suggestive form that looks like the original Cardy formula for an effective CFT. If we make the identifications
$$48  \pi R E_C \leftrightarrow {(d-2)c_{dS}} \qquad 2 \pi R E_{CFT} \leftrightarrow (d-2)L_0 \eqno(20)$$
then the extensive part of the energy is given by $(2 \pi R/(d-2))E_E=L_0+c_{dS}/24$.
The plus sign in the above formula is the modification required in the Euclidean CFT case dual to de Sitter space. Its origin is the sign change in metric in eq. (11)
when one goes from AdS to dS space.
(Another interpretation of the sign change is to take the negative Casimir energy to imply a negative central charge, $c<0$. This
is not surprising since we know that the CFT dual to $dS_d$ is not unitary.) Using eq. (20) the modified Cardy--Verlinde formula for the entropy becomes
$$S=2 \pi \sqrt{{c_{dS} \over 6}{(L_0+{c_{dS} \over 24})}} \eqno(21)$$
We find that $dS_d$ is described by a thermal state of the boundary (Euclidean) CFT with $c_{dS}=3 L^{d-2}/G_d$. This picture is very similar to the one that describes
the $AdS$ black hole in the boundary CFT. Both the $AdS_d$ black hole and pure $dS_d$ are described by a thermal state of a conformal boundary theory. Eq. (21) can be
considered as a modified Cardy formula.
It is quite surprising to find that the entropy of the Euclidean CFT on the boundary of de Sitter space is
given by this formula which is known to apply to two dimensional unitary CFTs. However, in this case the CFT lives in more
than two dimensions and is not unitary.

\medskip
\centerline{\bf 4. The Effective CFT and Asymptotically de Sitter Spaces}
\medskip

If $d$--dimensional de Sitter space is described by a $(d-1)$--dimensional Euclidean CFT on the boundary what would be the description of an asymptotically de Sitter space with a black hole?
This space does not have the conformal symmetry of the pure de Sitter space due to the existence of the black hole. Thus it must be described by a boundary theory
which is not conformal. For example, if the de Sitter space corresponds to a fixed point of the boundary theory then the asymptotically de Sitter space would live away from the fixed point.
In addition, for $M>0$,
the temperature of the asymptotically de Sitter space (or the cosmological horizon) is lower than the Hawking temperature of the black hole. As a result,
this space is unstable due to Hawking radiation from the black hole. As time increases the black hole mass $M$ decreases due to Hawking radiation. Finally
after the black hole completely evaporates and all the radiation crosses the cosmological horizon we reach a stable
state which is pure de Sitter. In the dS/CFT correspondence time evolution in the bulk is dual to the RG flow (from the IR to the UV) on the boundary theory[\INF,\BBM]. Therefore,
we expect to describe Hawking radiation by this RG flow from the IR to the UV, i.e. by
integrating in UV degrees of freedom on the boundary theory. The asymptotically de Sitter space is described by a nonconformal field
boundary theory with a monotonic $C$--function. As time passes the black hole evaporates and $C$ increases due to the RG flow. Finally the black hole
disappears and we reach a pure de Sitter space. On the boundary theory $C$ reaches the (conformal) UV fixed point with $C=c_{dS}$ which is the end of the RG flow.

The central charge of $dS_d$ can be obtained from the Casimir energy of the boundary theory by the identification in eq. (20). On the other hand, the Cardy--Verlinde formula is
valid for asymptotically de Sitter spaces as well. On the other hand, the Cardy--Verlinde formula is valid for asymptotically de Sitter spaces as well.
By the same reasoning we define the $C$--function of the nonconformal boundary theory which describes the asymptotically de Sitter space (with a black hole)
by making the identifications
$$48  \pi R E_C \leftrightarrow {(d-2)C} \qquad 2 \pi R E_{BND} \leftrightarrow (d-2){\tilde L_0} \eqno(22)$$
Here we have called the total energy $E_{BND}$ (since the theory is not conformal) and defined the $C$--function for the theory.
Then the extensive part of the energy is given by $(2 \pi R/(d-2))E_E={\tilde L_0}+C/24$.
The modified Cardy--Verlinde formula for the entropy becomes
$$S=2 \pi \sqrt{{C \over 6}{({\tilde L_0}+{C \over 24})}} \eqno(23)$$
It is quite surprising to see such a formula for the entropy of a theory which is not even conformal (in addition to being not unitary). However, we note that the conformal symmetry
of the boundary theory
is broken by the nonzero black hole mass $M$ (or $R<L$) and by the Hawking radiation which means $M$ is time dependent. On the boundary this translates into a scale dependent
radius $R$ for the volume of $S^{d-2}$.

From eq. (24) we find that
$$C={3 R^{d-3}L \over G_d}=c_{dS} \left({R^{d-3} \over L^{d-3}}\right) \eqno(24)$$
For nonzero black hole mass $M$, $R<L$ and therefore $C<c_{dS}$. This is simply a restatement of the fact that for any nonzero $M$ the cosmological horizon
is smaller than the horizon for the pure de Sitter case. The asymptotically de Sitter space corresponds to a nonconformal theory on the boundary since we are
away from the fixed point. The $C$--function defined above satisfies the two requirements of the c--theorem. First, it is a monotonic function of the scale or energy of
the boundary theory, i.e. it increases as the theory flows from the IR to the UV. As the black hole evaporates, $M$ decreases and $E_{BND}$ increases since $E_{BND} \sim -M$.
In addition, the radius of the cosmological horizon, $R$ (which can be written as $R \sim L-aM$ in any dimension) increases which means that $C$ increases with the energy.
The second requirement is satisfied because $C$ becomes equal to the central charge when the theory becomes conformal. From eq. (24) we see that
$C \to c_{dS}$ at the UV fixed point since then $R \to L$.

The relation between time evolution in the bulk of $dS_d$ and the RG flow on the boundary theory was demonstrated for the pure $dS_d$ case[\INF,\BBM]. Here we show that the same
relation continues to hold for the asymptotically de Sitter spaces[\DAN]. For this purpose it is better to use the cosmological coordinates defined by
$$\rho(t,r)=re^{-r/L} \qquad \qquad \tau(t,r)=t+{L \over 2}ln({r^2 \over L^2})-1) \eqno(25)$$
Then the metric for the asymptotically de Sitter space in eq. (2) becomes
$$ds^2=-N_{\tau}^2dt^2+h(d\rho+N_{\Sigma}d\tau)^2+r^2d\Omega_{d-2}^2 \eqno(26)$$
where
$$h={r^2 \over V} \left(1-{{r^2V^2} \over {L^2 V_0^2}} \right) \rho^{-2} \quad N_{\Sigma}={{1-{V^2 \over V_0^2}} \over {1-{r^2 V^2 \over {L^2 V_0^2}}}}{\rho \over L} \quad
N_{\tau}={\sqrt{V} \over {\sqrt{1-{{r^2V^2} \over {L^2 V_0^2}}}}} \eqno(27) $$
where $V_0=1-r^2/L^2$ and $V=1-r^2/L^2-2GM/r^{d-3}$. From the above metric we see that a time translation is equivalent to a rescaling of $\rho$ on the boundary. Therefore
time evolution is dual to an RG flow in the boundary theory even for asymptotically de Sitter spaces with a black hole.

We can now interpret the dynamical instability (in bulk time) in the asymptotically de Sitter space due to Hawking radiation from the black hole as an RG flow from the IR to
the UV on the boundary theory. Hawking radiation from the black hole decreases $M$ and therefore increases $R$ and $C$.
Above we saw that time evolution in the bulk corresponds to a scale transformation on the boundary theory.
As time passes in the bulk the black hole Hawking radiates and its mass gets smaller. However, as $M$ decreases the energy of the asymptotically de Sitter
space and therefore the energy of the boundary theory increases. ($E_{CFT}$ is negative and increases with decreasing $M$. For $dS_d$ $E_{CFT}=0$.) Thus,
the description of Hawking radiation on the boundary is an RG flow from the IR to the UV. Hawking radiation from the black hole
looks like integrating in UV degrees of freedom on the boundary theory which is the reason behind the monotonic rise of $C$.
At the end of black hole evaporation (and after all the radiation crosses the cosmological horizon) $M=0$ and $R=L$ so we reach de Sitter space.
From the boundary point of view, this is described by an RG flow with an increasing $C$--function which ends at the UV fixed (conformal) point with $C=c_{dS}$.

We see that, (for a given $\Lambda$ or $L$) $C$ is bounded from above by $c_{dS}$ which corresponds to a UV fixed point describing pure de Sitter space.
On the other hand, $C$ is also bounded from below by the fact that the smallest cosmological horizon is obtained by including the largest possible
(Nariai) black hole in the space [\NAR]. In this case, the cosmological horizon is at $R^2=[(d-3)/(d-1)]L^2$ so the minimal value of $C$ is $C_{min}=[(d-3)/(d-1)]^{(d-3)/2}c_{dS}$.
There is no bulk state which gives a lower entropy for thwe cosmological horizon; therefore there is no boundary theory with a smaller value of the $C$--function.
However, the Nariai black hole is not a stable configuration. A small
perturbation (such as decreasing the black hole mass) destabilizes the solution. When the black hole horizon becomes slightly smaller than the cosmological horizon, the black hole
starts radiating and gets smaller and smaller (until it completely evaporates). We conclude that the Nariai black hole corresponds to an IR (UV unstable) fixed point on the
boundary theory.

As an example, consider a black hole of mass $M$ in $dS_4$ which is described by a boundary theory with a $C$--function given by eq. (24).
As the black hole radiates $M$ decreases, but $R$ and $C$ increase. When the black hole completely evaporates, $M \to 0$ and $R \to L$. Then
$$C={3RL \over G_d}  \to {3 L^2 \over G_d}=c_{dS} \eqno(28)$$
The boundary theory becomes conformal since this is the UV fixed point of the RG. The lower bound on $C$ which arises from the Nariai black hole is
$C_{min}=3^{-1/2}c_{dS}$.

Asymptotically de Sitter space with $d=3$ is a special case since there is only one (cosmological) horizon even for $M>0$. In this case, $M$ does not describe
a black hole but a point mass with neither a horizon nor Hawking radiation. From eq. (24) for $C$ we see that $C=3L/G_d=c_{dS}$ for any $M$. Therefore,
asymptotically $dS_3$ space with any mass $M$ is described by a CFT at a fixed point of the boundary theory. Different masses correspond to different $L_0$ as given by eq. (20);
i.e. different excitations of the same CFT. In this case the lack of RG running on the boundary theory corresponds to the absence of Hawking radiation in the bulk. Note that in
this case the boundary theory is $1+1$ dimensional and conformal; therefore the existence of a Cardy--like formula for the entropy is less puzzling.


The Euclidean field theory on $I^+$ describes the cosmological horizon in de Sitter space (with or without a
black hole). In particular the Cardy--Velinde formula gives the entropy associated with the
cosmological horizon.
It seems more difficult to describe the black hole horizon and the entropy associated with it by using the Cardy--Verlinde formula.
If we insist on using the Cardy--Verlinde formula for the black hole entropy, we need a theory that lives on a boundary with a radius equal to that of the black hole horizon, $R_{bh}$.
Certainly this theory does not live on $I^+$ since it has a larger radius $R_{cos}$ in order to describe the cosmological horizon.
In any case the boundary theory on $I^+$ has a different entropy and temperature that those
required to describe the black hole. The only alternative is to have another theory on $I^-$ with radius $R_{bh}$. Then through the Cardy--Verlinde formula
this theory will automatically provide the correct
black hole entropy and temperature in an identical way to the cosmological horizon described above. The possibility of two different theories on the two boundaries $I^+$ and
$I^-$ is only possible because of the existence of the black hole. In pure de Sitter space lightlike curves connect the antipodal points on $I^+$ and $I^-$ and the
correlator between these point are singular. As a result, the two CFT are copies of each other and there is only one independent CFT on one of the boundaries. The
situation is different for the de Sitter black hole which has a very different Penrose diagram[\GH]. In this case, no such identification between the two boundaries seems to exist.
Thus, it seems that we may be able to put two different theories on the two boundaries.

\medskip
\centerline{\bf 5. Conclusions and Discussion}
\medskip

In this paper we gave a derivation of the Cardy--Verlinde entropy formula for the dS/CFT correspondence using the metric, the asymptotic conformal symmetry and the
Bekenstein--Hawking entropy in the bulk. This is certainly not a derivation of the Bekenstein--Hawking entropy. However,
if we believe that the area law is a fundamental feature of a true theory of quantum gravity, it would imply the Cardy--Verlinde formula for the dual field theory on $S^d$.
As we saw the Cardy--Verlinde formula can be put in a form very similar to the original Cardy formula by simple identifications.
When the bulk geometry is de Sitter (asymptotically de Sitter) the boundary theory is conformal (not conformal). It is quite surprising to find this formula for de Sitter
spaces since the boundary CFT lives in more than two dimensions and is not unitary whereas the original Cardy formula was derived for two dimensional unitary CFTs.
On the other hand, it is even more puzzling to find that the formula is valid for nonunitary and nonconformal boundary theories in more than two dimensions as we learn from
the asymptotically de Sitter spaces.

We also described asymptotically de Sitter spaces with a black hole by a nonconformal theory on the boundary with a $C$--function, $C<c_{dS}$ which is fixed by the Casimir
energy. We showed that $C$ is a nondecreasing function of the energy on the boundary. In this description, Hawking radiation from the black hole
can be seen as an RG flow from the IR to the UV in the boundary theory. Since, as the black hole evaporates  $C$ increases this looks like integrating in degrees of
freedom in the boundary theory. At the end of black hole evaporation (and after all the radiation crosses the cosmological horizon), we reach de Sitter
space which corresponds to the UV fixed point of the boundary theory with $C=c_{dS}$, i.e. a CFT. In addition, due to the existence of a maximum size black hole in
de Sitter space, $C$ is also bounded from below. The Nariai black hole
corresponds to an IR (unstable UV) fixed point of the boundary theory. For asymptotically de Sitter space in $d=3$ we find that $C=c_{dS}$ for any $M$. Thus there is no RG
run on the boundary which agrees with the fact that there are no black holes which radiate in $dS_3$.

The description of asymptotically de Sitter spaces by the dS/CFT correspondence is somewhat disappointing since the CFT lives on the future (or past) boundary which is way
beyond the cosmological horizon and not accessible to an observer. In our opinion, a more satisfying description would describe the space by degrees of freedom that live
on the horizon and are accessible to an observer. There is such a description for black holes in terms of strings with rescaled tensions that live on the stretched horizon[\LEN-\UNI].
A similar description can also be given for pure de Sitter space[\DES]. It seems that the same kind of description can be generalized for the asymptotically de Sitter space.
Since both descriptions have the same number of degrees of freedom, i.e. the same entropy, there must be a (probably nonlocal) mapping between them.
It would be interesting to find the connection between these two descriptions of de Sitter space.

\bigskip
\centerline{\bf Acknowledgements}

I would like to thank Raphael Bousso and Lenny Susskind for useful discussions.

\vfill

\refout

\end
\bye